\documentstyle[art12,a4]{article}
\begin{document}
\font\big=cmbx10 scaled\magstep1
\newcommand{\beq}{\begin{equation}}
\newcommand{\eeq}{\end{equation}}
\newcommand{\beqa}{\begin{eqnarray}}
\newcommand{\eeqa}{\end{eqnarray}}
\newcommand{\fig}{Fig.\,~}
\newcommand{\emline}{\vspace{.4cm}}
 \begin{titlepage}
  \begin{flushright}
   USITP-91-21\\
   November 1991\\
   Revised May 1992\\
  \end{flushright}
  \parindent 0pt
  \null
  \vskip 2cm
  {\Large
    \begin{center}
     The Probability for Formation of Collapsing
     Textures and Texture Knots.
    \end{center}
  }
   \vskip 1.5cm
   \begin{center}
    Stefan {\AA}minneborg\\
    \vskip .3cm
    Department of Physics\\
    Stockholm University \\
    Vanadisv\"agen 9\\
    S--113 46 Stockholm, Sweden\\
   \end{center}
   \vskip 1.5cm
   \begin{abstract}
  Global textures are interesting because they are promising candidates for
  seeds in the structure formation of the universe.
  The important configurations are those who will be able to collapse.
  The type of configurations that I will consider are half knots
  and true knots.
  I define a half knot as one that covers more than a half-$S^3$,
  and a true knot as a configuration that covers the whole of $S^3$.
  Configurations that are  half knots will be able to collapse,
  a new argument for this collapse criterion is given in this paper.

  I will also give some critical remarks on
  the method of using the total winding number
  as a criterion for identifying a knot.
  I propose another more direct method
  and make use of it when I estimate the probability for
  formation of both true and half knots by numerical simulations.
  In three dimensions  the probability for formation of
  a true knot is found to be $p_{knot}=0.003$
  and the probability for formation
  of a half knot is found to be $p_{half}=0.066$.
  Taking into account the chance of annihilation  the
  probability for collapse is found to be
  in the interval $p_{coll} \in [0.01,0.04]$.

  In the skyrmion picture of hadrons the investigations of
  formation of true knots are important, and some implications of my result
  on the predictions of  a "topological" theory of baryon production
  in jets are discussed.
   \end{abstract}
   \end{titlepage}

\section{Introduction}

The existence of a global symmetry, spontaneously broken in such a way
that the vacuum manifold is isomorphic to $S^3$,
seems to provide a promising explanation of the large-scale structure in the
universe \cite{first,struc}.

The simple but quite general idea
 \cite{first} is to assume that there exists a four component real scalar
field $\phi_a$ with trivial vacuum  above a critical temperature $T_c$,
and a degenerate vacuum at low temperatures  with the action
\beq\label{genact}
S=\int d^4x \sqrt{-g}(\frac{1}{2} \partial_\mu \phi_a
\partial^\mu \phi_a - \frac{\lambda}{4}(\phi_a  \phi_a - \eta^2)^2).
\eeq
Here $g$ is the determinant of the metric and $\lambda$ is a coupling
constant.
The symmetry-breaking scale $\eta$ is assumed to be around the
GUT-scale ($\approx 10^{16}\: GeV$).
This mechanism is possible since the effective potential
is temperature dependent in this way for a theory with the action
(\ref{genact}) at low energies,
it is also found
that $T_c\sim \eta$  \cite{breaking}.

In this scenario  pseudo-topological defects can be formed by
the Kibble mechanism \cite{kibble}.
The vacuum manifold is a $3$-sphere, an $S^3$.
We are here assuming that no inflation has
taken place. In the context of inflation there is also the
possibility of texture production but now in a different way
and with different features \cite{inflation}. The main mechanism
will then be quantum fluctuations of the inflaton field instead of
thermal fluctuations, although interesting this will not
be further discussed in this paper.

Some of the formed defects
will collapse \cite{prok} and produce density perturbations \cite{first}.
A numerical simulation of the evolution of textures
has been made by Spergel et al.  \cite{sim}, and some promising
results concerning the structure formation are found in \cite{struc}.
The textures will give an imprint on the cosmic background radiation
that is near the limits of observation \cite{cbr}.
The details of the collapse of a topological texture and the gravitational
effects have also been investigated \cite{grav}.
\emline

Let us consider, for the sake of generalisation,
 a space with $d$ spatial dimensions and in it a $d+1$ component
real scalar field $\phi_a$  whose
vacuum manifold after the phase transition is $S^d$.
For length scales larger than the inverse mass of the radial
"Higgs" mode $m_r^{-1}=(\lambda \eta^2)^{-1}$
the dynamics of the field can be described by the nonlinear $\sigma$
model \cite{first}.
Thus we have the simple action
\beq\label{lin}
S=\int d^dx \sqrt{-g}
\frac{1}{2} \partial_\mu \phi_a \partial^\mu \phi_a,
\eeq
 with the constraint $\phi_a \phi_a =\eta^2$.

After the symmetry breaking the field will be uncorrelated on
length scales larger than the horizon.
It seems  that the dynamics of the theory is such that
the correlation length always can be assumed to be of the same order
as the horizon length \cite{first,sim},
e.g. the spherically symmetric collapse occurs at the speed of light.

On scales smaller than the correlation length
 the field has to vary in such a way that it minimizes
the energy, the integral of the square of the gradient has to be
as small as possible.
In regions with a size of the same order as the correlation length
we can thus tell what the values of the field are.
This is given by the values of the field at the boundary and the demand
that the field has to be continous and sweep out as small
area (or hyper-area) as possible.
The winding number of a region
is defined by the fraction of $S^d$ that is swept out by the field
inside this region.

Since the field is uncorrelated on scales larger than the horizon
there can exist  regions in $R^d$ with a
diameter larger than the horizon length where the
field winds around $S^d$. With this is meant that all values in
$S^d$ are attained by the field in the region concerned, I
will call this a true knot. Notice that the total winding number
of such a configuration is larger than one.
True knots will collapse as is seen by Derrick's theorem \cite{derrick}.
More interesting though, is that true knots are not the only type
of configurations that can collapse. I will show for a simplified case
that configurations covering more than a half-$S^3$,
( they are called half knots), also want to shrink.
\emline

The probability of forming true texture knots
is of interest in other cases beside the cosmological texture model.
Consider e.g.  the model of
Ellis and Kowalski for "topological"
production of baryons in jets \cite{ellis1,ellis2},
this phenomenological model is inspired by the
skyrmion picture of baryons.
Here the correlation length is fixed (approximately 1 fm), and the defects
are assumed to be stabilized by the Skyrme term.
The global chiral symmetry is spontaneously broken at low temperatures
but assumed to be restored in the jet fragmentation process.
At low energies the chiral $\overline{q}q$ condensate field $U$
is a mapping $U:R^3\rightarrow SU(N)$, where N is the number of flavours.
After the phase transition which takes place in the jet fragmentation
region when it cools down it is thus possible that defects can be formed
by the Kibble mechanism.

If the chiral $\overline{q}q$
condensate field $U$ covers all of its target space on a
region of $R^3$ a baryon (or anti-baryon) is said to exist at that location.
In the simple two flavour case
$U:R^3\rightarrow SU(2)\sim S^3$, so my results
in this paper will be relevant here. This will be further
discussed in section \ref{baryon}.
\emline

In this paper I will focus on the initial conditions for the
structure formation, the number density of collapsing textures
and true texture knots.
I will thus adopt the basic ideas of \cite{prok,sriv}, that is to
study the probability for formation of the required type of defect
in a suitable small region.
In order to take into account boundary effects I will examine
the probability of annihilation.
I will throughout this paper only treat the case of a flat
Friedmann universe,
e.g. I assume $\Omega=1$.

A difficult problem one faces is to determine the
characteristics of a texture that is destined for collapse.
After that we must decide on the best ways to search
for this features.
The mostly used method of deciding when a configuration is representing
a collapsing texture is to look at the total winding number.
I will in this paper give some critical remarks on this method and
 propose an alternative method.

\section{True texture knots}

First I will  discuss the probability for formation of a true knot.

If we are given the boundary condition that the field at infinity
is constant, the problem is  topological since it is then possible
to compactify our $R^d$ to $S^d$.
The homotopy group of mappings from $S^d$ to $S^d$ is $\pi_d=Z$.
The total winding number is thus always an integer.

It is however very important to remember that in general there does not
exist any natural boundary conditions, this has also been emphasized
by  Prokopec and Srivastava  \cite{prok,sriv}.
In the lack of boundary conditions the problem is not  topological
in character since the homotopy group $\pi_d$ of $R^d$ is trivial.
We can therefore no longer use the total winding number for the
identification of a knot.

The relevance still of topological textures
is due to the fact that textures in three dimensions are unstable
against collapse. This will give the problem a "pseudo-topological"
character in the sense that the texture will evolve and after
a while can resemble a topological texture. We can then make use
of our knowledge of the dynamical properties of a collapsing
topological texture,
many investigations have been made about
its metric and the effects on surrounding matter among
other things \cite{grav}.

\subsection{Topological textures}
\label{topo}

As a first exercise I will calculate the probability for formation of
a topological texture knot, i.e. a configuration in an $S^d$ space
representating a winding number exactly equal to one.
This has of course been done before \cite{prok,ellis1} but
I will do it in a slightly different
way and I also hope that the procedure may shed some light
over the discussion about half knots.

We make a triangulation of $S^d$ with regular
d-simplices, the minimum number of simplices needed is $d+2$.
In \fig 1 this kind of triangulation is shown for the case of
$2$ and $3$ dimensions.

We let the distance between two neighbouring vertices be
equal to the correlation length.
At each vertex in the triangulation
the field is assumed to attain, uncorrelated with all the other vertices,
a randomly chosen value on $S^d$. Of course
each value of the field on $S^d$ has to be equally probable.

Since we demand the correlation length to be equal to the
length of the edges of the
simplices, the field on the inside of a simplex is restricted to the
interior of the
spherical polyhedron on $S^d$ that is spanned by the field values
at the simplex' vertices.
(This is however not completely true, see the remark in the next section.)

Knots are configurations such that
the field covers all of $S^d$,
all possible values of the field are attained and in the right order.
Such objects are often called  "hedgehogs".

We have a "hedgehog" on $S^d$ when the orientations of
the polyhedrons are all equal \cite{ellis1}.
The probability for positive orientation
is equal to that for negative. The probabilities are
 not correlated between the different simplices.
This can be seen from the fact that the number of simplices
are equal to the number of vertices with randomly chosen field values.
As we have $d+2$ simplices the probability that all the
spherical polyhedrons have the same orientation is $(1/2)^{(d+1)}$.

This means that the probability for forming a knot in an $S^d$ space
is $(1/2)^{(d+1)}$.

\subsection{Nontopological textures}

We now turn to the much trickier problem of estimating
the probability for formation of a non-topological texture knot
$p_{knot}$.

One could as Srivastava \cite{sriv} try
 to estimate the probability for
a configuration to be such that a region in $R^d$ of suitable size
 is compactifiable to $S^d$,
and then multiply this with the probability for having a knot in $S^d$.

Srivastava argues that if we want a configuration such that a region
is compactifiable we need a constant field on the boundary of the region.
He has calculated the probability for this occurence in a discretised
model.
His results implies,
in two dimensions $p_{knot}=8 \cdot 10^{-6}$, and in three dimensions
$p_{knot}=6 \cdot 10^{-9}$ \cite{sriv}.
However we do not have to be that restrictive and these
numbers seem to be much too small.
This has also been noticed by Borrill et al. \cite{borril}.

We can relax the demand that the field has to be constant at the
boundary, it is a sufficient but not necessary condition for having
a compactifiable region.
I will try to illustrate this in the case $d=2$.
By Srivastavas arguments we would need a simple closed curve in $R^2$
such that the image of it on $S^2$ is a point.
However also if the image of one simple closed curve in $R^2$
is a simple and unclosed curve
 on $S^2$, the region inside the simple closed curve
in $R^2$ will be compactifiable to $S^2$.
This is to say that we can make a sphere of the region inside the curve
by glueing the boundaries together (\fig 2).

Now it is not an easy task to calculate the probability for such
occurrences, so we need some other approach.
To begin with we must triangulate a large enough region in $R^d$ in
a suitable way. The distance between the vertices of each simplex
ought to be the correlation length.
In three dimensions it is unfortunately not possible to
make a triangulation with equal distances between all neighbouring
vertices, so we have to relax our demand in that case.
The region must be large enough for a knot to be able to
 form, so the diameter of the region has to be around
three correlation lengths.

We choose the direction of the field randomly
at each vertex, we assign a point on $S^d$ to each vertex.
We consider the area (or hyper-area)
of the spherical polyhedron spanned by the points on $S^d$
corresponding to the vertices of one simplex.
We define the absolute value of the winding number
of the simplex as this area (or hyper-area) divided by the area
(or hyper-area) of $S^d$ itself.
The sign of the winding number
is given by the orientation of the spherical polyhedron.

Of course it is not always the case that the field in reality must
attain all the values inside the spherical polyhedron described, but for
adjacent polyhedrons that does not matter, the net number
will become correct when we add together their winding numbers.
On the other hand, when it comes to
 the boundary of the region we really would need knowledge of the
situation outside the region in order to get the correct result.
Because of this the total winding number will be slightly overestimated.

If we add together all the winding numbers of the simplices
we get the total winding number of the whole region
and would imagine that we have a knot if the absolute value
of the winding number is larger than one.
This strategy is used by Prokopec \cite{prok}.

Now we have to be very careful.
The concept of total winding number is not a useful one since $\pi_d$
 of $R^d$ is trivial.
The condition that the total winding number is larger than one is surely
necessary but it is not sufficient for concluding
that we have a true knot.
It is possible that a part of
a spherical polyhedron with the same orientation appears more than once,
so we can get a total winding number larger than one but be left with
holes on the sphere,  values of the field that are not
attained at any point in the region.
I illustrate this with
 a two-dimensional example in \fig 3, here the
field only takes values in the lower half of $S^2$ despite that
the total winding number is found to be one.

\subsection{Numerical simulations of true knots}
\label{numtru}

To tackle the problem of identifying a true knot
we try a different approach.
In order to decide whether a configuration is a true knot or
not, we need to explicitly test each value of the field
and look if it is  attained somewhere in the region.
This is of course impossible, we have to restrict ourselves to
a finite number of test points uniformly distributed on the sphere.
We test one point on $S^d$ after another.
The algorithm for deciding whether a point is contained in a spherical
polyhedron or not can be found in \cite{ellis1}.
We  sum over all polyhedrons and check whether the point under test
gives a net contribution.

We take a set of $N$ test points uniformly distributed on the sphere,
and for each configuration  test whether all of the test points are
contained in the configuration or not.
We test  $n$ configurations on the region used,
and count how many times $n_{pass}$ all
the test points pass the test, this is done for different numbers $N$.
The output is the relative number density $p=n_{pass}/n$.

In order to imitate that we test all points on the sphere we have to
incorporate a trick.
The number $p$ that we get out of the computation can be
interpreted as an estimate for
the probability $p(N)$ that all the test vectors will pass the test
for a given $N$.
This is of course larger than the probability of creating
a true knot $p_{knot}$, because we have not tested all points
on the sphere.
The statistical error is calculated using
$(\Delta p)^2=p\: (1-p)/(4 n)$.

The probability $p(N)$ is theoretically
\beq
p(N)=p_{knot} + \int_{0}^{1}(1-x)^{N}f(x) dx,
\eeq
where $f(x)dx$ is the probability to get a texture that miss to
cover a fraction $x$ of the sphere, and $(1-x)^N$ is the probability
that all the test points miss this fraction of the sphere.
By partial integrations we can write $p(N)$ as
\beq
p(N)=p_{knot} + \sum_{i=0}^{\infty} \frac{N!}{(N+i+1)!}f^{(i)}(0)
\eeq
If we expand this in terms of $\frac{1}{N}$ we get for the
first three terms,
\beq
p(N)\approx p_{knot} + \frac{1}{N}
f(0)+(\frac{1}{N})^2 ( f'(0)-f(0)).
\eeq
To get $p_{knot}$ we fit this curve to our results from the
simulations, as a bonus we also get $f(0)$ and $f'(0)$.
With this method we also get a very accurate error estimation for
$p_{knot}$.

I have used this method in order to estimate the probability $p_{knot}$
for formation of a true knot in
$2$ and $3$ dimensions.
I will later ( in section \ref{half})
use this method also for estimating the probability for
formation of a half knot.

The region used in $2$ dimensions
is depicted in \fig 4. It consists of $12$ vertices and $13$
regular triangles.
The results of the simulations in $2$ dimensions are shown in \fig 5.
We find $p_{knot}=0.0197\pm0.0004$.

The region used in $3$ dimensions is a part of a body centered
cubic lattice, this lattice can be divided into irregular tetrahedrons
as is depicted in \fig 6.
This triangulation has been used by others in the context of monopoles
 \cite{cubic}.
Each vertex is common to $24$ tetrahedrons.
My region consists of a "middle" tetrahedron and all the tetrahedrons
that have any vertex in common with this "middle" tetrahedron.
This will give us a region consisting of $32$ vertices and
$71$ tetrahedrons.

The results of the simulations in $3$ dimensions are shown in \fig 7.
We find $p_{knot}=(2.9\pm 0.1)\cdot 10^{-3}$.

Because I ignored the boundary effects  this number is a slight
overestimation. We will however not have the risk of annihilation of
two true knots, the knots collapse before the pair is inside the horizon.

\section{"Topological" production of baryons}
\label{baryon}

As mentioned in the Introduction a case  where the results
in this paper concerning true knots
are of interest is  the model of
"topological" production  of baryons in jets.
The true knots in this theory are called skyrmions and are
identified with baryons and anti-baryons.

In the original treatment of the model, Ellis and Kowalski
\cite{ellis1,ellis2}
did not use quotation marks around topological.
They introduced a hypothetical fourth dimension in order
to compactify $R^3$
in to $S^3$ and thus be able to define the winding number.
Although unphysical this fourth dimension certainly has physical
implications \cite{sriv},
without its introduction we are concerned with
mappings $R^3\rightarrow S^3$ and they are trivial.

By their construction they found that the probability for formation
of a skyrmion in one simplex is $1/16$ since
this is the probability
for formation of a topological texture in $3$ dimensions
 (see sec. \ref{topo}).

Without any boundary conditions the number should rather be
$0.003$ as  found above.
The exactness of my result can  be discussed since
I neglected boundary effects and the simplices
were not regular.
Nevertheless this number should be quite a good estimate
for the real probability,
furthermore it is an overestimation.

A very important feature of the problem of baryon production in
jets is that the regions of hadronization are quite small.
Outside the jet the field is confined to be constant.
Thus we do have some fixed boundary conditions and the
number $0.003$ will only be appropriate for large hadronization
regions and then well inside the jet.
The size of the hadronization regions will typically increase with
energy.

According to my analysis above, the numbers of Ellis and Kowalski
on the baryon multiplicity at high energies have to be decreased.
For small hadronization regions their treatment serves as a good
approximation, though.

Also the dependence on sphericity will be affected, the baryon
multiplicity will increase less with increased sphericity
than in their treatment.

Their baryon momentum distribution has to be corrected
because the partons momenta are not uniformly distributed in
the region and the probability for baryon creation is larger
near the boundary.

\section{Collapsing cosmic textures}

In the case of cosmological textures  the possibility of later
collapse of a texture is the important circumstance.
When a half knot, a region in which the field
covers more than a half-$S^3$
(note that this is not equivalent with  covering more
than half of $S^3$),
comes inside the horizon it is energetically
most preferable for it to collapse. This will be discussed in more detail
later. The collapsing texture
will give rise to density perturbations \cite{struc}.
When the size of the texture shrinks to zero it will eventually unwind
itself \cite{first}.

It is common \cite{prok,prle} to call a region with a winding number
greater than one half, a half knot. This has a drawback, it is
possible that there exist
regions with a total winding number more than one half
where the field does not cover a half-$S^3$.
I have already discussed this kind of problems in the context of true knots.

The arguments for the collapse criterion directly use the fact that
the field covers more than one half-$S^3$ \cite{first,prok},
 rather than the total winding being greater than $1/2$. This has
however not been emphazised before as far as I know,
I also illustrate this with a new argument in section \ref{colcri}.
\emline

Now there are many ways to explore the pattern of the textures
and the number of collapses as a function of time.

One strategy is to study for a certain small region of minimal
size, the probability for forming a texture with winding number
large enough.
This in turn can be made in several ways.
Two main philosophies are the discrete \cite{sriv}
 and the continous  \cite{prok} models.
In the discrete
model one restricts the field to $d+2$ values uniformly distributed
on $S^d$. Even if the continous model is harder to handle I think it
is to be preferred due to its greater accuracy
as is illustrated by Prokopec \cite{prok}.

Because the texture knots unwind and disappear, one would imagine
that the number of collapses is constant in a flat universe.
The situation looks the same at all times,
the so called scaling solution \cite{first}.
That this is true has been indicated in actual simulations on
large grids by Spergel et al. \cite{sim}.
This holds only for a flat universe, it is of course not true for a closed
or open universe.
Their result is that the number of unwinding
textures per horizon volume per unit time is constant and equal to
$0.04$ with a relative error of about $50\%$.
This would mean that the probability
of creating in a minimal region a configuration which is
destined for collapse is equal to $0.04$.

Another strategy is to consider large lattices and randomly
distribute field values on the points of the lattice,
afterwards the number of textures that may collapse is counted.
This is done by Borrill et al. \cite{borril}
using the discrete approach, and by Leese and Prokopec
 \cite{prle} using the
continuous approach. Here one great difficulty is that
the knots are quite unlocalised so the results are hard to
interpret.
Borrill et al. use the concept of partial winding in order
to solve the localisation problem, they do not use the
concept of total winding number at all.
They find the relative number of collapsing textures to be $0.035$,
it is said to be an overestimation
and having a relative error of about $50\%$.

The probability for formation of a collapsing texture
is not equal to the probability for formation of a half knot, there is
also the chance that some of the half knots will annihilate each
other before they start to collapse \cite{borril}.
I will later in this paper in more detail discuss the probability for
creation of half knots and their annihilation.
I will then find a result on the relative number of collapsing
textures consistent with the above mentioned.

\subsection{The collapse criterion.}
\label{colcri}

If we want to examine the problem of the collapse criterion
we have to study the non-linear $\sigma$ model in more detail.

In order to simplify the discussion of the collapse we look at the
"spherically symmetric" ansatz
\beq
\phi({\bf r})=\eta (\cos \chi(r,t), \hat{\bf r} \sin \chi(r,t) ),
\eeq
where $\hat{\bf r}$ is the unit vector in the direction of
${\bf r}$. This ansatz satisfies the constraint
$\phi_a \phi_a = \eta^2$.
I use quotation marks because this ansatz is symmetric only under
simultaneous rotation in both space and field space.
We consider only spherically symmetric collapse, this is however
quite general as I will argue later.

With flat Robertson-Walker metric
\beq
ds^2=R(t)^2(dt^2-dr^2-r^2(d\theta^2+sin^2\theta d\phi^2)),
\eeq
the action (\ref{lin}) for this ansatz becomes
\beq
S=4\pi \eta^2\int^{\infty}_{-\infty} dt\: R^2(t)
\int^{\infty}_{0} dr\: r^2((\dot{\chi})^2-(\chi')^2-2\frac{sin^2\chi}{r^2}),
\eeq
where dot stands for partial derivation with respect to the
conformal time $t$, and  prime stands for partial derivation
with respect to the radial coordinate $r$.
If we vary $S$  and put $\delta S=0$
we get the equation of motion
\beq\label{whole}
\ddot{\chi}
=\chi''+ \frac{2}{r}\chi' -\frac{\sin 2\chi}{r^2}.
\eeq
We have here taken $R(t)$ to be constant which is a good approximation
when the horizon is large.
The equation we get if we put $\chi(r,t)=f(r)$ has been investigated
by Iwasaki and Ohyama
\cite{japan},
\beq\label{stat}
f''+ \frac{2}{r}f' - \frac{\sin 2f}{r^2}=0.
\eeq
They  found that that the only finite boundary conditions at zero
and infinity that can be upheld by the solutions $f(r)$ of
this equation are
\beq\label{solu}
f (0)=n \pi, \; f(r\rightarrow \infty)=n \pi \pm \frac{\pi}{2}.
\eeq
An important observation is that the solutions of (\ref{stat})
thus always correspond to textures covering exactly a half-$S^3$.

Let us now take a solution $f(r)$ of (\ref{stat})
with boundary conditions  $f (0)=0$ and
$f(r \rightarrow \infty)= \frac{\pi}{2}$, the existence of such an
$f$ is guaranteed by the proof by Iwasaki and Ohyama
 and that is all that we need.
We take as initial conditions $\chi(r,t_0)=a f(r)$ and $\dot{\chi}(r,t_0)=0$,
 where $a$ is a constant.
If we put this into (\ref{whole}) we get
$\ddot{\chi}(r,t_0)>0$ when $a>1$ and $\ddot{\chi}(r,t_0)<0$ when $a<1$.

This is because  when $0<2f<\pi$, for $a>1$ we have $\sin 2af<a\sin 2f$
and for $a>1$,
$\sin 2af>a\sin 2f$.

What we have found is that when the special "spherically symmetric" texture
we are considering  covers more than a half-sphere
it tends to collapse, but if it covers  less than
a half-sphere it tends to expand.
Furthermore since there are no other solutions to (\ref{stat}) than
(\ref{solu}) $\ddot{\chi}$ can never change sign again, so the
collapse or the expansion will go on for ever.
Although this holds only for a one parameter set of initial
configurations it indicates that the critical
property of a configuration that determines the collapse is
whether it covers more than a half-$S^3$ or not.
In this case that coincides with the winding number being greater than
$1/2$ but my assertion is that the property of covering more than
a half-$S^3$ is the crucial one.

Recent investigations indicates that the critical winding number for the
collapse of a "spherical" texture is somewhat greater than $0.5$.
Two different inconsistent results exist, $0.74 \pm 0.03$
\cite{borr92} and $0.602 \pm 0.003$ \cite{prokoct91}.
These are the numbers  for a flat universe,
the results for the other cases can be found in the references.
The main reason for the critical winding number being greater than
$0.5$ is that the half knot in practise is not infinite, opposite
to my implicit  assumption in the analysis above.

In general we are not dealing with "spherically symmetric"
configurations.
It is  possible that the "spherical" solutions acts as attractors
in the same way that the self-similar solution appears to do,
although this ought to be further examined.
There is a claim that knots become spherical in the collapse,
presumably because other modes than the spherical are radiated away
\cite{cbr}.

Because of the above arguments I will in the following
assume that collapse will take place for configurations covering more than
a half-sphere, a half-$S^3$, and they will be called half knot textures.
Observe here that according to this definition a true knot is also
a half knot.

\subsection{Texture half knots.}
\label{half}

I will now discuss the probability for formation of a half knot texture.

We can make a good estimate for the probability for
formation of a half knot by a simplification.

Let us consider the kind of triangulation
that is shown in \fig 8 for $2$
and $3$ dimensions. The general procedure in d dimensions
is to start with a regular d-simplex and add its centre.

To use this kind of triangulation  is of course
not entirely correct since
the distance between the centre and one outer vertex is not
equal to the distance between two neighbouring outer vertices.
If we take the distance between the centre and the other vertices
to equal the correlation length, our construction means that we
are neglecting non-correlations occuring between the  simplices'
outer vertices. The field on the simplices is thus not entirely
given by the field on the vertices.
It is hard to predict wether this will give an enhancement
 or a suppression of the number of half knots,
Another way to interpret the situation would be to say that we have
a space that is curved in such a way that the distances between
nearby vertices are equal.

The construction described above is essentially the same as the
triangulation of $S^d$ considered before.
The difference is that we now do not treat the polyhedron
consisting of only the outer vertices as a simplex on its own.
This fact enables us to conclude
 immediately that the probability for forming a half knot
is $(\frac{1}{2})^{(d+1)}$.
The field on the hypothetical simplex consisting of
only the outer vertices can not
span more than a half-$S^d$, so if we would have had an
$S^d$-texture for a given
configuration, the $d+1$ polyhedrons that we are considering must wind over
more than a half-$S^d$.

In three dimensions we thus arrive to the conclusion that
  the probability for forming
a half knot is $1/16$, which should be
compared with $0.04$ and $0.035$,
the probability for formation of a texture
that will collapse according to Spergel et al. \cite{sim}
and Borrill et al. \cite{borril}, respectively.

Is the main reason for the difference between my
number and theirs the simpification  made when calculating the probability
for half knots?

The answer is no.
This will be shown in the next section.

\subsection{Numerical simulations of half knots}

In order to examine the properties of more appropiate
triangulations than the one described above
 I have made simulations similar to the ones in the calculation
of the probability for true knots.
There is  only one modification of the method used that is
needed. When a test vector ( test point) has not passed
the test I test the inversion of the test vector.
If, for all vectors, the vector itself or its
inversion is contained in the configuration we have
a half knot, because the configuration then contains a half-$S^d$.

The region used in $3$ dimensions
is based on one vertex of the body centered cubic lattice
described in section \ref{numtru}
 and consists of all the $24$ tetrahedrons that
contain this special vertex. We thus have $14$ vertices around the
central vertex.

I have used the same type of analysis as in the case of true knots
and the results are shown in \fig 9.
We get $p_{half}=0.066\pm 0.001$, this is even greater than the number
$1/16$ found by the simplification above.

\subsection{Collapsing Textures}

If we want to know the probability for formation of a collapsing
texture we have to take into account the boundary effects we
have been neglecting before, we must consider the
chance of annhilation.

If we have a half knot in our region
the probability for formation of a half knot with a winding number
of opposite sign,
in a neighbouring region is approximately $p_{half}/2$.
This is if we neglect the fact
that the probabilities  are  correlated to each other.
The probability for forming a half knot in a region,
when we already have one in another region sharing one
of its faces, is slightly enhanced.

With $n_{reg}$ neighbouring regions  the probability for annihilation
of one formed half knot is
\beq
\frac{p_{half}}{2} \sum_{i=0}^{n_{reg}-1}(1-\frac{p_{half}}{2})^i .
\eeq
This can be approximated by $n_{reg} p_{half}/2$
since $p_{half}/2$
is much smaller than $1$.
We now can conclude that
 the probability for  formation of a collapsing texture is
\beq
 p_{coll}=p_{half}(1-\frac{n_{reg}}{2} p_{half}).
\eeq
Notice that this means that $p_{coll}$ can never be greater than
$\frac{1}{2n_{reg}}$.
In three dimensions, the lowest realistic number of neighbouring regions
is $12$, since the area of a sphere with radius equal to the horizon
length is approximately $12$ "horizon areas".
We can thus conclude that the number of collapsing
textures per horizon volume per expansion time probably
does not exceed $1/24$.

Since the region in the triangulation I considered has $24$ faces
we have in this case $ p_{coll}=0.014\pm 0.001$, the errors are only
statistical.

My triangulation is not a universal one and other ones could be
more appropriate. We can try to incorporate this uncertainty
by taking quite conservative error limits;
$p_{half}\in [0.06,0.07]$ and $n_{reg}\in [12,24]$.
This will give us $p_{coll}\in [0.01,0.04]$.

\section{Conclusions}

We have seen that we need to be careful when we try to identify
 texture knots and half knots.
I have pointed out that we can not use the concept of
total winding number when identifying a true knot.
Instead I have used the method of explicity test whether all
values of the $S^3$ field are attained in a certain region or not.
With a modification of this method I have also been able to
estimate the probability for formation of half knots.
A half knot is here defined as a configuration that covers more than
a half-$S^3$. These configurations are interesting  because this feature
indicates collapse as shown in section \ref{colcri}.
However, two adjacent half knots
with different orientations
 will annihilate each other, and in that case there will
not be any collapse.
Taking this into account I found the probability for collapse
to be $p_{coll}=0.014\pm0.001$ for a special triangulation on
a cubic lattice also considered by others in the context of monopoles
\cite{cubic}.

In general I found $p_{coll}$ to be in the interval,
$p_{coll}\in [0.01,0.04]$.
This is consistent with the results of Borrill et al. $0.035$ \cite{borril}
and Spergel et al. $0.04$ \cite{sim} .
In both these works they assume there relative errors to be around
$50\%$, furthermore Borrill et al. claim that their number is
an overestimation.
In addition, I have found that of these collapsing
textures approximately $1/10$ are true knots. This may be important
 as true knots resemble topological textures, with gravitational
properties and impact on surrounding matter that are better understood
than those of general half knots.

  {\large
    \begin{center}
     Acknowledgement
    \end{center}
  }
I would like to thank Lars Bergstr\"om and Per Ernstr\"om
for helpful discussions and  suggestions.

\newpage

  {\large
    \begin{center}
     Figure Captions
    \end{center}
  }

\fig 1 : (a) Triangulation of  $S^2$, the vertices numbered $0$ are
all to be identified with each other. Observe that this means that
we get a tetrahedron if we make use of a third dimension and
fold up the triangles containing the vertex $0$.

 (b)  Triangulation of  $S^3$. It consists of a tetrahedron $1234$
with a tetrahedron attached to each of its faces, giving a total
of five tetrahedrons. Also here  the vertices numbered $0$ are
all to be identified with each other. If we here make use of a
fourth dimension we get a regular elementary $4$-simplex by folding
"up" the tetrahedrons containing the vertex $0$.

\emline

\fig 2 : A region (a) in $R^2$ compactifiable to $S^2$.
The curve $C(AB)$ is mapped on a certain curve $Im(C(AB))$ on $S^2$ (b),
and the curve $BA$ is mapped on the very same curve,
$Im(C(AB))=Im(C(BA))$.
\emline

\fig 3 : A region (a) in $R^2$ which is mapped on the lower half of
$S^2$ (b) two times giving a total winding number of $1$.
\emline

\fig 4 : The triangulation used in the simulation of true texture knots
in $2$ dimensions.
\emline

\fig 5 : Result of a simulation of the probability for formation of
 true texture knots in $2$ dimensions.
The solid line shows the best fit.
$p_{knot}=0.020\pm0.001$,
$f(0)=2.0\pm0.4$,
$f(0)-f'(0)=80\pm40$,
$\chi^2=1.6$.
\emline

\fig 6 : One of the tetrahedrons on the body centered cubic lattice
that is used when constructing
the triangulation used in the simulation of true texture knots
in $3$ dimensions.
\emline

\fig 7 : Result of a simulation of the probability for formation of
 true texture knots in $3$ dimensions.
The solid line shows the best fit.
$p_{knot}=(2.9\pm 0.2)\cdot 10^{-3}$,
$f(0)=7\pm1$,
$f(0)-f'(0)=(2.5\pm1)\cdot 10^3$,
$\chi^2=1.04$.
\emline

\fig 8 : Triangulations in (a) $2$  and (b) $3$ dimensions
used in a simple estimation of the probability
for formation of a half knot.
The vertex $0$ in (a) is situated at the centre of the
triangle $123$.
Similarly the vertex $0$ in (b) is situated at the centre of the
tetrahedron $1234$.
Observe the similarity between these  and the triangulations
shown in \fig 1.

\emline

\fig 9 : Result of a simulation of the probability for formation of
 half knots in $3$ dimensions.
The solid line shows the best fit.
$p_{knot}=0.066\pm0.001$,
$f(0)=65\pm7$,
$f(0)-f'(0)=(39\pm8)\cdot 10^3$,
$\chi^2=0.66$.

\end{document}